\documentclass[10pt,conference]{IEEEtran}
\IEEEoverridecommandlockouts

\usepackage{booktabs}
\usepackage{tabularx}
\usepackage{array}
\usepackage{fontspec}
\usepackage{microtype}
\usepackage{url}
\makeatletter
\g@addto@macro\UrlBreaks{\do\a\do\b\do\c\do\d\do\e\do\f\do\g\do\h\do\i\do\j\do\k\do\l\do\m\do\n\do\o\do\p\do\q\do\r\do\s\do\t\do\u\do\v\do\w\do\x\do\y\do\z\do\A\do\B\do\C\do\D\do\E\do\F\do\G\do\H\do\I\do\J\do\K\do\L\do\M\do\N\do\O\do\P\do\Q\do\R\do\S\do\T\do\U\do\V\do\W\do\X\do\Y\do\Z\do\0\do\1\do\2\do\3\do\4\do\5\do\6\do\7\do\8\do\9}
\makeatother
\usepackage{cite}
\usepackage[hidelinks]{hyperref}
\usepackage{xcolor}
\usepackage{tikz}

\newcolumntype{Y}{>{\raggedright\arraybackslash}X}
\newcolumntype{L}[1]{>{\raggedright\arraybackslash}p{#1}}
\newcommand{\stage}[1]{\textnormal{\textsf{#1}}}

\newcommand{\status}[1]{\textsf{#1}}

\title{Localizing Post-Wire Semantic Changes in MCP Agent Frameworks}

\author{\IEEEauthorblockN{Aditi Patodiya}
\IEEEauthorblockA{Independent Researcher\\
aditi.patodia31@gmail.com}\thanks{This work has been submitted to the IEEE for possible publication. Copyright may be transferred without notice, after which this version may no longer be accessible.}}

\begin{document}
\maketitle

\begin{abstract}
Valid Model Context Protocol (MCP) messages do not guarantee that an agent
framework preserves the distinctions downstream software needs. We present a
differential testing method that follows a fixed tool result through each
framework's public interfaces and checks explicit consumer requirements.
Applied to 18 designed fixtures in four pinned Python integrations, the method
identifies 13 unique fixture--task divergences involving structured values,
declared errors, and rich content. Google ADK satisfies every primary contract
in its observed path; the other integrations show interface-specific changes
or an execution failure. Whether a change matters depends on the consumer:
treating absent optional fields as equivalent to null explains most of OpenAI's
strict rich-content failures. In an exploratory replay, parsing JSON text
recovers more structured values but also returns incorrect values and values
for fields absent at the source. Documented settings help in conflicting-output
stress cases without restoring a separate structured slot. Fault challenges
expose an oracle weakness and test its repair on fresh cases. The study provides
reproducible, interface-specific evidence from controlled cases, not production
failure rates or model-behavior measurements. Its practical implication is that
post-wire regression tests need to specify both the information required and
how the consumer reads it.
\end{abstract}


\section{Introduction}

Protocol validity does not guarantee that downstream software receives the
semantics a server emitted. A Model Context Protocol (MCP)
\texttt{tools/call} response may contain an
ordered sequence of typed content blocks, separate \texttt{structuredContent},
an \texttt{isError} flag, and result metadata~\cite{mcp-tools-2025}. After the
official client decodes that response, an agent framework can retain those fields
in an application object, flatten them into text, split them between model and
auxiliary channels, or serialize a subset for a model provider. Each carrier can
be legal in its own application programming interface (API) while no longer
supporting a downstream consumer's task.

The official MCP roadmap identifies this result-handling boundary as an open
problem. It says
result handling falls short because equivalent output can appear in multiple
forms and a server developer cannot know which form a client presents to the
model; the maintainers aim to standardize a clearer contract~\cite{mcp-roadmap-2026}.
Individual framework issue records likewise describe structured-content,
metadata, and duplication behaviors~\cite{langchain-issue-283,microsoft-issue-7866,
google-discussion-3893}. Yet an issue report isolates one implementation event.
It does not reveal where an unchanged result diverges across native framework
carriers, whether the difference matters to a typed consumer, or whether the
measurement procedure catches previously unseen semantic faults.

Existing MCP research addresses complementary questions. Benchmarks evaluate
agent task completion and tool orchestration~\cite{guo2026mcpagentbench,
dong2026etom}; security studies characterize the ecosystem and protocol-level
attack surfaces~\cite{hou2026landscape,li2026confuseddeputy}; and AgentCheck
injects tool-response faults to evaluate end-to-end agent behavior and
mitigations~\cite{mazumder2026agentcheck}. Official conformance tests stop at the
protocol boundary~\cite{mcp-conformance-2026}, while shared agent specifications
enable controlled cross-runtime evaluation~\cite{amini2026openagentspec}.
Our question is narrower: after a valid result has crossed the wire, at which
public framework carrier does a task-relevant distinction become altered,
unavailable, or unobservable?

We answer one research question:

\begin{quote}
\textbf{RQ.} Where do task-relevant MCP result semantics become altered,
unavailable, or unobservable in four pinned agent-framework paths?
\end{quote}

The contribution is a testing method that separates three questions
conflated in interoperability checks: is the message valid, does the required
information survive, and can a particular consumer access it? Each integration
has its own observation path, reflecting the interfaces it actually exposes.
Executable contracts specify the distinctions a consumer needs at each interface;
separately frozen fault challenges test whether the measurement instrument can
detect their loss. We retain both the instrument's initial failure and its repair.

The controlled four-framework study demonstrates why these distinctions matter.
Structured data may survive as text yet be unavailable to a consumer expecting a
separate field. A reader that parses the text can recover more values, but its
choice of channel also determines how it handles conflicting outputs. Our
exploratory replay measures that tradeoff, including incorrect values and values
returned for fields absent at the source. The accompanying artifact preserves
the captured evidence, failed and unobservable cases, and deterministic
reconstruction of the results. Together, the method and study provide a way to
turn a broad interoperability concern into a testable interface requirement.

\section{Background and Related Work}

\subsection{MCP result handling}

MCP connects hosts, clients, and servers through JSON-RPC. In revision
2025-11-25, a tool result can carry text, image, audio, embedded-resource, and
resource-link blocks; a separate structured object; a declared error flag; and
metadata~\cite{mcp-tools-2025}. The specification recommends a text counterpart
for structured output for backward compatibility, but it does not require every
host to expose one common post-decoding representation. The newer roadmap's call
for a clear result contract is first-party evidence that client projection is an
open interoperability concern~\cite{mcp-roadmap-2026}. SEP-2200 proposed guidance
for result visibility, but was closed without merging~\cite{mcp-sep2200-2026}.
We treat it as prior articulation of the problem, not a normative requirement
against which these integrations fail.
The open SEP-3279 proposes negotiated, audience-specific outputs in place of
\texttt{content} and \texttt{structuredContent}~\cite{mcp-sep3279-2026}.
It targets a prospective December 2026 revision; neither adoption nor deployment
is established. Our pinned result format provides a baseline for testing such
transitions.

Hou \emph{et al.} characterize the MCP landscape, security threats, and research
directions~\cite{hou2026landscape}; Li \emph{et al.} study a confused-deputy
attack against MCP~\cite{li2026confuseddeputy}. MCP-AgentBench and ETOM assess
real-world task performance and orchestration~\cite{guo2026mcpagentbench,
dong2026etom}. Those outcomes combine planning, model behavior, tool execution,
and result handling. In contrast, our unit begins with an already-produced result
and ends before remote-provider processing.
SGVEF-LOOP evaluates MCP-agent behavior through coverage-guided metamorphic
exploration~\cite{liu2026sgvef}; we localize deterministic carrier changes before
model behavior.

Yang \emph{et al.} analyze clause implementation across MCP software development
kits (SDKs) using a
language-independent representation and study compatibility-abusing
attacks~\cite{yang2026compatibility}. Our measurements concern application
framework carriers after decoding, not missing SDK clauses or exploitability.

AgentCheck records agent/tool interactions, perturbs tool responses with 12 fault
types, and evaluates behavioral mitigation~\cite{mazumder2026agentcheck}. We use
faults for a different purpose: to challenge a deterministic measurement oracle,
not to claim generic fault-injection novelty or assess agent recovery. Mehan
audits whether completed MCP failure results expose enough information for
deterministic recovery decisions~\cite{mehan2026actionability}. Our T3 boundary
is narrower still: it asks only whether a declared-error distinction survives a
framework carrier and makes no recovery-actionability claim. Open Agent
Specification supports controlled comparison by executing shared agent and
workflow designs across four runtimes and three benchmarks, measuring accuracy,
latency, and execution behavior~\cite{amini2026openagentspec}. We instead fix one
tool result and inspect its native carriers. These complementary units separate
agent-level execution differences from post-decoding result projection.

\subsection{Testing boundary and novelty}

The official conformance suite validates captured MCP messages~\cite{mcp-conformance-2026}.
The \texttt{mcp-statecheck} tool adds generated stateful sequences, normalized
wire-trace comparison, shrinking, and replay~\cite{mcp-statecheck-2026}.
Those lifecycle checks complement our fixed-result, post-decoding boundary.
Same-input differential testing and semantics-aware API comparison show why
syntactic equality is insufficient~\cite{godefroid2020differential,
ma2026apidiffer}. Metamorphic relations further provide executable expectations
when no single output string is canonical~\cite{chen2018metamorphic}. We combine
these principles at a boundary not covered by wire conformance: one valid result
is observed through native application, provider-neutral, and locally serialized
provider carriers. The contribution lies in connecting framework-specific
observations to explicit consumer requirements, with fault challenges that test
the measurement itself. The four-framework study demonstrates this connection;
it does not depend on claiming that the individual framework behaviors were
previously unknown.

\section{Method}

\subsection{Task-relative post-wire fidelity}

Post-wire fidelity is defined here by what a consumer must distinguish. For
example, a renderer may need a resource's URI as well as its text, while a
structured-value reader may need to distinguish an absent field from an explicit
null. We call these required distinctions semantic atoms and bind them to the
part of the result that the consumer reads, its compartment.

Formally, a source result $r$ comprises typed semantic atoms $A(r)$: presence versus
null, structured type and value, content-block kind and order, exact text,
decoded binary payload, URI, MIME type, descriptors, annotations, error state,
and metadata. A stage projector $\pi_{f,s}$ reads only prospectively allowlisted
fields for framework $f$ at stage $s$. It preserves named compartments and
distinguishes absence from null; unrestricted recursive search is prohibited.

A deterministic consumer task binds a subset of atoms to a compartment and a
decision rule. A task passes when all of its required typed distinctions remain
recoverable from that compartment. A representation change can therefore pass:
a data URL may preserve media bytes and MIME type, for example. Conversely, a
structured value found only in an auxiliary artifact cannot satisfy a model-client
contract that requires a distinct structured slot.

Five tasks make the construct executable. T1 requires exact content value,
cardinality, and source order. T2 distinguishes missing, null, empty, and populated
structured state, recovers the required value and type, and preserves the
relation between structured output and parseable text (neither, text only,
structure only, equivalent, or conflicting).

T3 requires an explicit success or
error status: declared false maps to continue, declared true or a corresponding
error exception to abort, and an absent or null status to unknown. This is
deliberately stricter than the MCP schema's default of success when
\texttt{isError} is unset~\cite{mcp-result-schema-2025}: a consumer requiring an
explicit signal can distinguish declared success from an undeclared outcome.

T4 dispatches
by rich-content kind and recovers payload, identity, descriptors, annotations,
and block-level metadata. T5 separately enforces an application policy for
result-level metadata. The primary family contracts are T2 for structured values, T3
for error state, and T4 for rich content. T1 and T5 are guard tasks and are
reported separately; they cannot increase the primary-divergence count.

These contracts describe consumer requirements, not extra obligations imposed
on every MCP host. In particular, a T2 failure can mean that an object survives
only as JSON text. A consumer that parses that text might still recover a value,
whereas a consumer bound to a distinct structured slot cannot. Similarly, a
text-only renderer need not retain all T4 resource descriptors. We use the
stronger contracts to make the tested interface explicit; changing that interface
changes the interpretation of a pass. Protocol conformance, strict-contract
preservation, and application-specific recoverability are separate claims.

\subsection{Concrete downstream dependencies}

Public consumer implementations show why these distinctions can matter to
software, while also showing why no single projection policy is universal.
In a post-outcome source audit, we inspected version-pinned implementations and
tests released before the study freeze. Table~\ref{tab:consumers} connects their
observable software dependencies to the task families. These examples were
selected for construct interpretation; they are neither a consumer prevalence
sample nor a new evaluation cohort.

PydanticAI's MCP result mapper extracts structured values and its error path
selects retry, failure, or exception behavior~\cite{pydanticai-mcp-consumer-2026}.
VS Code dispatches resource links using their kind, URI, and MIME type, including
URI-directed reads for image resources~\cite{vscode-mcp-consumer-2025}.
MCP Inspector's existing component tests require a structured-output section
and check that expanding a resource invokes the read callback with its
URI~\cite{inspector-mcp-consumer-2026}. These are concrete uses of typed values,
error signals, and resource identity outside the measurement instrument.

Their differences are equally important. PydanticAI can unwrap a singleton
\texttt{result} object, while VS Code intentionally serializes structured data
as model-facing JSON text. Neither behavior establishes our dedicated-slot T2
contract as a universal requirement. Nor does URI-driven rendering require every
descriptor checked by the complete T4 atom. The sources anchor practical uses of
the distinctions; the exact compartment and preservation policies remain
explicit, authored requirements. We inspected these upstream tests but did not
claim to rerun them or use their outputs as independent gold labels.

\begin{table}[t]
\caption{External consumer dependencies and the limits of their support.}
\label{tab:consumers}
\centering
\footnotesize
\setlength{\tabcolsep}{3pt}
\begin{tabularx}{\columnwidth}{@{}L{0.23\columnwidth}Y Y@{}}
\toprule
Consumer & Software dependency & Not established \\
\midrule
PydanticAI 2.40.0 & Structured extraction; error-controlled retry/failure & Universal structured compartment or original wrapper \\
VS Code 1.104.0 & URI/MIME-directed resource reads and typed rendering & Preservation of every rich descriptor; structured-slot necessity \\
MCP Inspector 2.5.0 & Separate structured display; URI passed to resource read callback & Model-provider contract or model-quality effect \\
\bottomrule
\end{tabularx}
\end{table}

\subsection{Executable comparison and localization}

The comparison has two outputs: a task result and a localization boundary. For
each source, framework, and supported configuration, the harness records the
public carrier before scoring it. It then applies the allowlisted projector,
selects the task's compartment, and evaluates presence, type, and value against
the source requirements. Captured task outcomes remain separate from execution
failure and absence of observation. Repetition collapse occurs only after the
carrier states and every applicable task readout agree.

Localization is deliberately conservative. A passing observation followed by
an unmet observation bounds a change between those carriers; it does not prove
which internal function caused it. If the official decoded result is not
observable, a difference first seen at the application carrier can be localized
only to the wire-to-application interval. If two observations share one object,
different consumer readouts identify a compartment boundary without implying an
object transformation. This makes the method useful across frameworks whose
public APIs expose different amounts of their internal execution.

A failing contract also produces an interpretable diagnostic. For structured
state, the diagnostic distinguishes an absent slot from a mismatched type or
value. For rich content, it identifies missing, added, or altered required atoms,
including kind, URI, payload, and order. Such diagnostics support small regression tests: retain the
source fixture, name the public seam, and assert the required downstream
decision. They avoid making a provider call the only way to demonstrate an
adapter-level change.

\subsection{Framework-specific observation lanes}

The source value \stage{W0} is the canonical MCP response. \stage{D1} is the
same-invocation object decoded by the official SDK when a supported public seam
exists. \stage{A2} is the framework's public application carrier. \stage{M3} is
the public provider-neutral payload accepted by the model integration.
\stage{H4} is the provider-SDK request body captured at an injected local
transport; it makes no claim about TLS bytes, remote normalization, model input,
or model behavior.

The provider requests follow two schemas. Google's Gemini
\path{functionResponse.response} object retains the MCP result fields;
OpenAI and Microsoft use Responses \path{function_call_output} carriers.
Both paths use strings or typed content parts, depending on the fixture, without
a dedicated structured slot. An \stage{H4} result thus reflects the provider interface as
well as the framework, rather than an adapter-only difference.

\begin{figure*}[t]
\centering
% Observation boundaries, not a universal five-stage transformation chain.
\begin{tikzpicture}[x=1pt,y=1pt,font=\footnotesize,
  carrier/.style={draw,rounded corners=2pt,minimum height=22pt,inner sep=3pt},
  flow/.style={->,line width=0.5pt},
  missing/.style={text=black!65,font=\scriptsize}]
\node[anchor=east] at (107,144) {Observation};
\node[align=center] at (140,144) {\stage{W0}\\Wire};
\node[align=center] at (219,144) {\stage{D1}\\SDK decode};
\node[align=center] at (306,144) {\stage{A2}\\Application};
\node[align=center] at (382,144) {\stage{M3}\\Model client};
\node[align=center] at (476,144) {\stage{H4}\\Provider request};

\node[anchor=east] at (107,107) {OpenAI Agents};
\node[carrier] (ow) at (140,107) {\stage{W0}};
\node[carrier] (od) at (219,107) {\stage{D1}};
\node[carrier] (oa) at (306,107) {\stage{A2}};
\node[carrier] (om) at (382,107) {\stage{M3}};
\node[carrier] (oh) at (476,107) {\stage{H4}};
\draw[flow] (ow) -- (od); \draw[flow] (od) -- (oa);
\draw[flow] (oa) -- (om); \draw[flow] (om) -- (oh);

\node[anchor=east,align=right] at (107,70) {LangChain\\MCP Adapters};
\node[carrier] (lw) at (140,70) {\stage{W0}};
\node[carrier] (ld) at (219,70) {\stage{D1}};
\node[carrier,minimum width=114pt,minimum height=30pt] (ls) at (344,70) {};
\node at (306,74) {\stage{A2}}; \node at (382,74) {\stage{M3}};
\node[font=\scriptsize] at (344,63) {shared-carrier views};
\node[missing,align=center] at (476,70) {not applicable};
\draw[flow] (lw) -- (ld); \draw[flow] (ld) -- (ls);
\node[font=\scriptsize,anchor=west] at (288,47) {C14: A2 fails; M3 not reached};

\node[anchor=east] at (107,19) {Google ADK};
\node[carrier] (gw) at (140,19) {\stage{W0}};
\node[missing,fill=white,inner sep=2pt] at (219,30) {PNO};
\node[carrier,minimum width=114pt,minimum height=30pt] (gs) at (344,19) {};
\node at (306,23) {\stage{A2}}; \node at (382,23) {\stage{M3}};
\node[font=\scriptsize] at (344,12) {shared-carrier views};
\node[carrier] (gh) at (476,19) {\stage{H4}};
\draw[flow,dashed] (gw) -- (gs); \draw[flow] (gs) -- (gh);

\node[anchor=east,align=right] at (107,-22) {Microsoft\\Agent Framework};
\node[carrier] (mw) at (140,-22) {\stage{W0}};
\node[missing,fill=white,inner sep=2pt] at (219,-11) {PNO};
\node[carrier,minimum width=114pt,minimum height=30pt] (ms) at (344,-22) {};
\node at (306,-18) {\stage{A2}}; \node at (382,-18) {\stage{M3}};
\node[font=\scriptsize] at (344,-29) {separate compartments};
\node[carrier] (mh) at (476,-22) {\stage{H4}};
\draw[flow,dashed] (mw) -- (ms); \draw[flow] (ms) -- (mh);
\end{tikzpicture}
\caption{Framework-specific observation lanes, not a universal execution chain.
Boxes enclose distinct captured carriers; paired \stage{A2}/\stage{M3} labels
inside one box denote views or compartments, with no transformation arrow
between them. Arrows bound observations, not internal causal functions. Dashed
arrows span an unobservable \stage{D1} seam (PNO: predeclared not observable).
LangChain has no applicable \stage{H4} path; its C14 failure is retained in
Table~\ref{tab:results}.}
\label{fig:lanes}
\end{figure*}
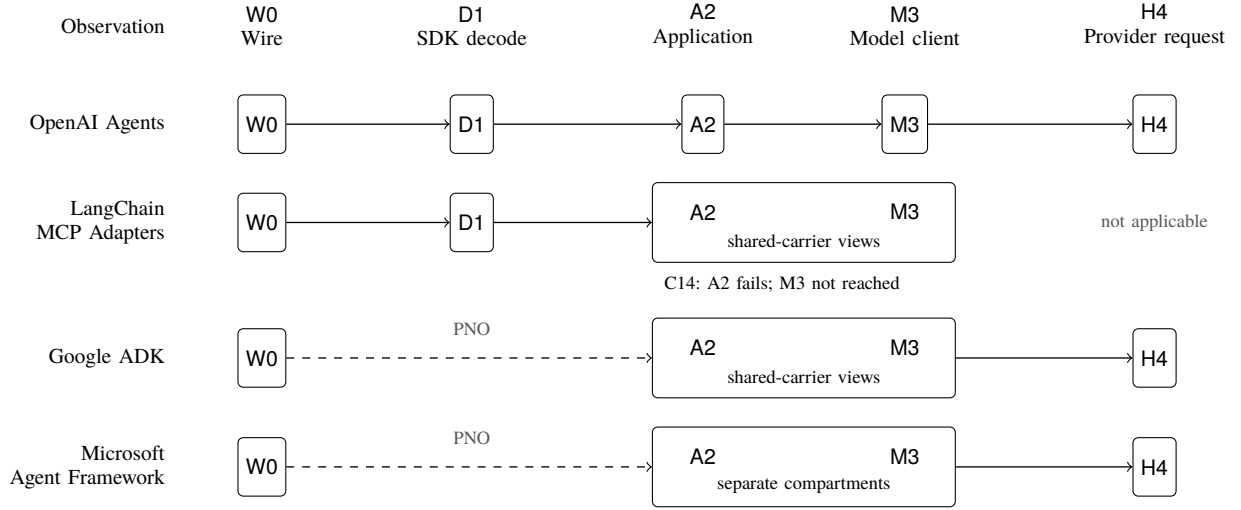

Fig.~\ref{fig:lanes} shows the actual study topology. OpenAI Agents exposes the
whole five-stage lane. Google ADK and Microsoft Agent Framework do not expose
\stage{D1} through the accepted public seam. LangChain has no intrinsic provider
path in the pinned adapter cohort, so \stage{H4} is not applicable. Google and
LangChain expose \stage{A2}/\stage{M3} views of a shared carrier; Microsoft
exposes separate compartments of one object. These observations distinguish
consumer scopes without implying object transformations or independent evidence.
Only OpenAI captures \stage{A2} and \stage{M3} as successive objects.

\subsection{Systems, fixtures, and execution}

The four independently maintained Python integrations in this purposive cohort
are OpenAI Agents 0.22.0, LangChain MCP Adapters
0.3.2 with LangChain 1.4.0, Google ADK 2.8.0, and Microsoft Agent Framework core
1.17.0 with its OpenAI integration 1.14.2. All paths use CPython 3.12.14, MCP
Python SDK 1.29.1, and protocol revision 2025-11-25. The versions and source
commits are pinned in the artifact~\cite{openai-agents-022,langchain-adapters-032,
google-adk-280,microsoft-af-117,mcp-python-1291}.
Revision 2025-11-25 was the installed SDK's \texttt{LATEST\_PROTOCOL\_VERSION}.
Among these exact releases and selected extras, only OpenAI admits SDK 2.1.1;
the other three require MCP below 2~\cite{openai-agents-022,langchain-adapters-032,
google-adk-280,microsoft-af-117}. This is a common supported baseline,
not a claim to cover the latest protocol.
The superseding 2026-07-28 revision is stateless, removes initialization, and
requires \texttt{resultType}, including an \texttt{input\_required} interim
variant~\cite{mcp-changelog-2026}. It retains the five content-block kinds and
\texttt{isError}, expands \texttt{structuredContent} to any JSON value, and treats
omitted \texttt{resultType} from earlier servers as complete~\cite{mcp-tools-2026,
mcp-changelog-2026}. Our result-shape measurements inform backward compatibility;
the newer lifecycle and interim results are untested.

Eighteen local fixtures cover three primary families: four ordinary structured
states (C01, C03, C05, C06), two error states (C07--C08), and eight rich-content
cases (C09--C16). Two structured stress cases (C02, C04) and two metadata-policy
cases (C17--C18) exercise guards but cannot establish the primary finding. Each
framework/fixture cell ran in three fresh processes. The admitted run contained
324 records: 216 primary/default; 36 configuration records (two prespecified
documented settings $\times$ six fixtures $\times$ three runs); and 72 direct
controls (four pinned environments $\times$ 18 fixtures $\times$ one run).
Direct controls bypass adapters and require \stage{W0}--official MCP
\texttt{ClientSession} \stage{D1} semantic identity; they are validity checks,
not outcome replicates. Repetitions are stability checks, not samples. A collapsed unit is
deterministic only when all three stage states and readouts agree; majority vote,
selective retry, and complete-case deletion are disallowed.

Table~\ref{tab:fixtures} summarizes these evidentiary roles. Values, identifiers,
payloads, and feature combinations were seed-derived and fixed before the
admitted run.

\begin{table}[t]
\caption{Fixture groups and evidentiary roles.}
\label{tab:fixtures}
\centering
\footnotesize
\begin{tabularx}{\columnwidth}{@{}L{0.16\columnwidth}L{0.12\columnwidth}Y L{0.14\columnwidth}@{}}
\toprule
Cases & Task & Contract distinction & Role \\
\midrule
C01, C03, C05, C06 & T2 & Missing, populated, falsy, and empty structured state & Primary \\
C07--C08 & T3 & Explicit false versus true error state & Primary \\
C09--C10 & T4 & Plain text versus typed embedded-text resource & Primary \\
C11--C16 & T4 & Link descriptors, heterogeneous order, image, audio, and blob identity & Primary \\
C02, C04 & T1/T2 & Structured-only and conflicting text/structure & Guard \\
C17--C18 & T5 & Result metadata absent versus present & Guard \\
\bottomrule
\end{tabularx}
\end{table}

The protocol, fixtures, projectors, tasks, dependency trees, order, and analysis
rules were hash-frozen before the admitted execution. Raw provider-client bodies,
exceptions, lifecycle traces, and provenance preceded derived scoring. The run
made no external model-provider request, attempted no guarded network access,
used no automatic retry, and cost USD~0.

\subsection{Instrument challenge and disclosed repair}

The original oracle passed a 48-case development contract, but that corpus was
written alongside the instrument and is not held out. We therefore built a
separate frozen validation with fresh identifiers, values, operators, and carrier
instances. Expected semantics were anchored to the official MCP schema and tools
specification snapshots. Twelve injected faults changed a task contract and 12
metamorphic controls changed representation without changing the contract. The
instrument had not been developed against these cases.

The first suite detected 11 of 12 faults and accepted all 12 controls. Its sole
miss swapped the text bodies of two embedded text resources: T1 noticed the order
change, but T4 did not because its rich signature omitted exact text. We preserved
the failed result and made one general repair: T4 now compares the complete
encoding-independent semantic atom, including exact text, payload, MIME, URI,
descriptors, annotations, and metadata. No other task or projector changed. A
second, separately frozen 24-case suite used new values and operators,
including a three-resource text rotation. It detected 12/12 faults and accepted
12/12 controls. These counts describe complete, constructed suites. No random
fault-generating population was specified, so we do not interpret them as
population accuracy estimates or attach sampling-based confidence intervals.

Post hoc, after both validation outcomes were known, we cross-scored the frozen
corpora without changing them or either instrument. V1 again detected 11/12
faults (missing HVF008 and HV2F008); v2 detected 12/12, and both accepted 12/12
controls. The corpora share no case identifier, input hash, or operator-family
label. Suite 002 therefore discriminates v1 from v2, although it was designed
after the T4 weakness was known. Eleven controls per suite test T1 and one T4;
none tests T2, T3, or T5.

Table~\ref{tab:validation} preserves the failed and successful challenges;
the development result is not accuracy evidence. Both fresh suites required
every fault and control to be classified correctly, valid pinned sources, no
execution error, and no hash drift.

Two deliberately simpler checks provide development-set baselines. A schema/wire
check accepts any schema-valid carrier and therefore detects none of the 36
semantics-changing but valid mutations. A channel-insensitive recursive check
finds familiar values anywhere in a carrier; it detects 24/36 mutations but
incorrectly rejects three of 12 lossless controls. These outcomes demonstrate
which distinctions the checks cannot express. They are not comparative
performance estimates because the same authored corpus informed the typed method.

\begin{table}[t]
\caption{Development checks, frozen gates, and post-hoc (PH) cross-scores on
constructed suites. Counts are finite-suite outcomes, not accuracy estimates
for a sampled fault population.}
\label{tab:validation}
\centering
\footnotesize
\setlength{\tabcolsep}{3pt}
\begin{tabular}{@{}lccc@{}}
\toprule
Method/corpus & Faults & Controls & Role/gate \\
\midrule
Schema/wire, development & 0/36 & 12/12 & baseline \\
Recursive, development & 24/36 & 9/12 & baseline \\
Typed v1, fresh 001 & 11/12 & 12/12 & fail \\
Typed v2, fresh 001 & 12/12 & 12/12 & PH \\
Typed v1, fresh 002 & 11/12 & 12/12 & PH \\
Typed v2, fresh 002 & 12/12 & 12/12 & pass \\
\bottomrule
\end{tabular}
\end{table}

\subsection{Fail-closed execution history}

Five earlier runs were excluded in full: four stopped during collection and one
failed admission after collection when its Microsoft H4 projector rejected valid
rich-block fields. Corrections to execution, reconstruction, and validation code
received new locks and full restarts; the fixtures, research question,
thresholds, and tasks did not change. Separately, a broad metadata search exposed
fragments from excluded runs. We recorded the incident, stopped raw inspection,
and did not interpret or use the exposed values. Correction scope was checked
before the admitted run. The audit trail preserves
each exclusion and correction; none of these runs contributes reported outcomes.

\subsection{Analysis}

The analysis includes every planned default-profile record. A captured task is
\status{pass} or \status{requirement unmet}; other states remain explicit as
\status{execution failed}, \status{not reached}, \status{not applicable}, or
\status{predeclared not observable}. A \emph{unique divergence} is one
fixture--primary-task pair for which, at the same observable stage, at least one
framework passes and another has a captured unmet requirement. Multiple stages
for that pair are manifestations of one divergence. The complete matrix is a
designed finite census, so we report exact counts without p-values or population
failure rates. The instrument-v2 reanalysis was locked after the validation
repair and reverified all source and raw-run hashes before computing results.

\subsection{Exploratory consumer-policy sensitivity}

To test whether strict T2 failures imply lost value recoverability, we added a
post-outcome sensitivity analysis on the existing C01--C06 records. This set
includes the four ordinary structured fixtures and the two original stress
cases. It adds no framework executions. Its question, native field paths,
decoder code, unit tests, and input inventory were locked before the new readouts
were computed, but the original outcomes were already known. It is therefore
exploratory analysis, not another held-out validation.

An independently implemented standard-library decoder supports two policies
without importing the study's projectors or oracle. A \emph{native-slot} reader
retrieves a structured field only from its
declared native path. A \emph{slot-or-JSON-text} reader prefers any present slot,
including null or empty values; if absent, it parses each entire native text
item as JSON. It rejects prose-wrapped, concatenated, or duplicate-key JSON,
requires all text items to parse, and returns ambiguity for unequal typed
candidates. It neither searches auxiliary compartments at model-facing stages
nor selects whichever substring matches the source.

The target is the source's structured-field presence and exact typed value.
Boolean, number, string, null, array, and object distinctions remain explicit;
list order matters, while object-key order does not. Correct absence counts as a
match. A value introduced when the source field was absent is recorded separately
from missing, wrong, ambiguous, or unparseable output. This target deliberately
omits T2's text/structure relation and uses a nested native-field absence policy:
for example, LangChain's null artifact has no \texttt{structured\_content} field.
The original projector instead preserves that null artifact as an observed null.
The policies are alternatives, not corrections silently substituted for T2.

We verified all 324 normalized-record hashes, selected 108 framework records and
24 direct controls, and collapsed three repetitions into 108 planned stage
positions: 96 captured and 12 LangChain H4 positions not applicable. All repeated
readouts agreed. The 24 direct controls matched under both policies. The reported
counts describe these designed positions, not sampled recovery probabilities.

\subsection{Rich-content optional-null sensitivity}

A post-hoc T4 sensitivity tests whether absent optional descriptors or metadata
containers need to remain distinct from explicit nulls. It was specified after the original outcomes were
known and selected atom differences had been inspected. The rule normalizes
source and observed atoms symmetrically, preserving populated annotations,
nested metadata entries, kind, payload, identity, and order. It uses the same
projectors and complete-atom comparison, with no framework-specific exception,
fixture exclusion, or change to T2 and T3. A further post-hoc recount compares
the full cohort and every single-framework omission under both policies.

\begin{table*}[!t]
\caption{Primary-task passes, listed in \stage{A2}, \stage{M3}, \stage{H4}
order. Each entry is pass/denominator. Square brackets group two views or
compartments of one carrier; bars separate carriers. Only OpenAI's first two
entries represent successive objects. Parentheses retain execution failure (EF) or not reached (NR).
A dash is not applicable.}
\label{tab:results}
\centering
\small
\setlength{\tabcolsep}{5pt}
\begin{tabularx}{\textwidth}{@{}L{0.23\textwidth}Y Y Y@{}}
\toprule
Framework path & Structured state/value (T2; $n=4$) & Declared error (T3; $n=2$) & Rich identity/order (T4; $n=8$) \\
\midrule
Google ADK & $[4/4;\,4/4] \mid 4/4$ & $[2/2;\,2/2] \mid 2/2$ & $[8/8;\,8/8] \mid 8/8$ \\
LangChain MCP Adapters & $[3/4;\,1/4] \mid {-}$ & $[2/2;\,2/2] \mid {-}$ & $[1/8\,(1\,\mathrm{EF});\,1/8\,(1\,\mathrm{NR})] \mid {-}$ \\
Microsoft Agent Framework & $[0/4;\,0/4] \mid 0/4$ & $[1/2;\,1/2] \mid 0/2$ & $[1/8;\,1/8] \mid 1/8$ \\
OpenAI Agents & $3/4 \mid 1/4 \mid 1/4$ & $2/2 \mid 0/2 \mid 0/2$ & $1/8 \mid 1/8 \mid 1/8$ \\
\bottomrule
\end{tabularx}
\end{table*}

\section{Results}

\subsection{Public interfaces determine localization}

In OpenAI's successive-object lane, the structured-value requirements for C03,
C05, and C06 and both
error-state requirements pass at \stage{A2} but become unmet at \stage{M3}.
LangChain's \stage{A2}/\stage{M3} differences instead occur between compartments
of a shared carrier; it has no provider-request path. Where \stage{D1} is
unobservable, failures first observed at \stage{A2} remain localized only to the
\stage{W0}--\stage{A2} interval. The scored bindings below exclude \stage{D1}.

All 216 default-profile records were admitted, and the 368 scored
fixture--framework--stage--task bindings contained no repetition disagreement.
Google, Microsoft, and OpenAI captured all 18 \stage{A2}, \stage{M3}, and
\stage{H4} carriers. LangChain captured 17 of 18 \stage{A2}/\stage{M3} carriers;
its C14 audio case failed consistently at \stage{A2}, leaving \stage{M3} not
reached. All 18 LangChain \stage{H4} positions remained not applicable. These
states were retained and never converted to task failures or removed.

The validated repair changed interpretation, not capture. Relative to instrument
v1, v2 leaves every carrier state unchanged but recognizes C13 and C14 as primary
rich-content divergences, increasing the unique primary count from 11 to 13 and
stage manifestations from 32 to 38. It
also withdraws a v1 directional \stage{H4} text/resource contrast: under the
complete T4 atom, OpenAI's C10 carrier no longer satisfies the full embedded-resource
contract. The workshop paper therefore reports the post-validation v2 analysis
only; it does not retain a favorable v1 contrast under the repaired construct.

\subsection{Preservation depends on the interface and contract}

Information retained for application code need not remain available at the
model-facing interface. Table~\ref{tab:results} shows this distinction across
the primary contracts. Google satisfies every designed primary contract across
three views of two distinct downstream carriers. OpenAI retains structured and error
information at \stage{A2} but not generally in \stage{M3}/\stage{H4}; LangChain's
application artifact supports three of four ordinary structured cases while its
model-client compartment supports one; Microsoft turns structured output into
ordinary text and therefore satisfies none of the four T2 cases. Rich-content
identity is the most selective contract: apart from the eight Google passes,
each other framework passes only the plain-text case at each available stage,
with the disclosed LangChain audio failure.

These differences affect all three primary families
(Table~\ref{tab:families}). Across them, 13 unique fixture--task divergences occur:
four structured, two error-state, and seven rich-content divergences. Observing
the same distinction at multiple stages produces 38 manifestations (12, 5, and
21 by family). The headline count is 13 distinct fixture--task pairs, not 13
independent defects. Seven additional
T1 fixture--task divergences occur in the primary fixture set. Stress and
result-metadata cases remain separate in the artifact; none enlarges the primary
count.

\begin{table}[!b]
\caption{Primary findings count fixture--task pairs once and stage observations
separately. Pairs need not correspond to independent defects.}
\label{tab:families}
\centering
\footnotesize
\begin{tabularx}{\columnwidth}{@{}Ycc@{}}
\toprule
Primary family & Unique fixture--task pairs & Stage manifestations \\
\midrule
Structured state/value (T2) & 4 & 12 \\
Declared error state (T3) & 2 & 5 \\
Rich identity/order (T4) & 7 & 21 \\
\midrule
Total & 13 & 38 \\
\bottomrule
\end{tabularx}
\end{table}

Under the strict contract, Google is the sole passing comparator for all seven
rich-content pairs. Complete-atom equality is the reference contract because it
preserves the distinction between an absent descriptor and an explicit null;
the alternative below describes consumers that do not need that distinction.
A leave-one-framework-out recount under the strict contract retains six unique divergences without Google,
all in structured state or declared errors; the remaining rich-content paths
still fail their contracts but no longer form pass/unmet contrasts. Removing
any other framework retains all 13 pairs. This dependence describes the selected
cohort, whose differing stage availability and designed fixtures do not support
a population ranking.

Optional-field nulls explain 18 of OpenAI's 21 unmet T4 positions. The post-hoc
check covers all 96 planned rich-content positions (86 captured), of which 54
have captured unmet outcomes under the strict contract. All 18 changes to pass
are in OpenAI's path: its score becomes 7/8 at every stage, with C13 still unmet.
Every other score is unchanged. The full cohort still yields 13
unique divergences and 38 manifestations, but OpenAI now joins Google as a
passing comparator for six rich-content pairs. Without Google, the optional-null
recount retains 12 of 13 unique pairs, compared with six under the strict policy.
Omitting Microsoft also leaves 12; omitting LangChain or OpenAI leaves all 13.
Thus both attribution and cohort sensitivity depend on the consumer's null
policy, even when the full-cohort count is stable. The artifact preserves
both readouts and every changed atom; Table~\ref{tab:results} retains the original
strict contract.

\subsection{Documented settings help under a different consumer policy}

Enabling documented structured-output handling did not restore a dedicated
structured slot at the tested model-facing interfaces. The prespecified
comparison paired defaults with OpenAI's
\texttt{use\_structured\_content=True} or LangChain's documented interceptor on
C01--C06. Neither setting rescued T2: all 18 matched outcomes were unchanged.
Twelve of 18 T1 guards moved from pass to unmet; six were unchanged, and none
improved. This exact task-level summary was computed post hoc after outcomes were
known; settings, fixtures, and endpoints were frozen before the run. These are
18 framework--fixture--stage pairs, not 18 independent deployments: LangChain
contributes six M3 pairs, and OpenAI contributes six M3 and six H4 pairs. The
comparison does not include the application-only artifact as an alternative
model-client channel.

The result concerns the strict T2 interface. A setting that serializes a
structured object into ordinary text can be useful to a JSON-parsing consumer
without restoring the dedicated slot. The T1 changes likewise identify changed
text/cardinality/order under the original renderer contract, not measured harm
to a large language model (LLM). Consequently, these settings should be assessed against both the
consumer's parsing policy and the text contract that the setting may replace.

For consumers that parse JSON text, however, the settings can be useful.
Table~\ref{tab:recovery} reports exact source matches under the two
consumer policies. OpenAI's setting increases JSON-text matches from 3/6 to 5/6
at both M3 and H4; LangChain's interceptor increases M3 matches from 4/6 to 5/6.
The gains occur in the two original stress fixtures, not the four ordinary
structured fixtures. Thus the settings can help a value consumer even though
the frozen strict T2 outcomes are unchanged.

Exact matches alone hide a consequential tradeoff. Across the 96 captured
framework--configuration--stage positions, JSON-text fallback raises matches
from 49 to 74 but returns five wrong values and nine values where the source
structured field was absent (Table~\ref{tab:recovery-outcomes}). The native-slot
reader has no wrong-value readouts, yet also returns two source-absent values:
OpenAI's application carrier exposes null for an absent field in both settings.
Neither policy is therefore uniformly fail-safe. The comparison shows how much
a consumer recovers alongside the kind of error it may return, not just how
often it produces a value. The counts include the original stress fixtures and
shared-carrier views.

\begin{table}[t]
\caption{Exploratory exact source matches per stage over C01--C06, including
correct absence. Grouped stages/configurations have identical counts; they are
not additional independent samples. LC H4 remains not applicable.}
\label{tab:recovery}
\centering
\footnotesize
\setlength{\tabcolsep}{3pt}
\begin{tabularx}{\columnwidth}{@{}Y L{0.19\columnwidth}cc@{}}
\toprule
Path/configuration & Stage(s) & Slot & +JSON \\
\midrule
Google/default & A2, M3, H4 & 6/6 & 6/6 \\
LangChain/both & A2 & 6/6 & 6/6 \\
LangChain/default & M3 & 1/6 & 4/6 \\
LangChain/interceptor & M3 & 1/6 & 5/6 \\
Microsoft/default & A2, M3 & 1/6 & 4/6 \\
Microsoft/default & H4 & 1/6 & 1/6 \\
OpenAI/both & A2 & 5/6 & 5/6 \\
OpenAI/default & M3, H4 & 1/6 & 3/6 \\
OpenAI/structured & M3, H4 & 1/6 & 5/6 \\
\bottomrule
\end{tabularx}
\end{table}

\begin{table}[t]
\caption{Complete exploratory readouts over the same 96 captured positions.
Source-absent means a returned value when the source structured field was
absent; it does not mean an LLM hallucination.}
\label{tab:recovery-outcomes}
\centering
\footnotesize
\begin{tabularx}{\columnwidth}{@{}Yrr@{}}
\toprule
Outcome & Native slot & Slot or JSON text \\
\midrule
Exact match (including correct absence) & 49 & 74 \\
Missing required value & 45 & 1 \\
Wrong type or value & 0 & 5 \\
Value at source-absent field & 2 & 9 \\
Ambiguous candidates & 0 & 3 \\
Unparseable text & 0 & 4 \\
\midrule
Total captured positions & 96 & 96 \\
\bottomrule
\end{tabularx}
\end{table}

The conflict fixture C04 makes the mechanism concrete. Its structured object
declares \texttt{decision: deny}, while its text counterpart declares
\texttt{decision: allow}. At OpenAI M3/H4, the default text reader returns the
wrong \texttt{allow}; the structured setting returns the source's \texttt{deny}.
At LangChain M3, the interceptor exposes both candidates, changing a wrong
\texttt{allow} into an explicit ambiguity. Neither path gains a dedicated
structured slot. These are deterministic readouts of synthetic fixture fields,
not observations of an agent authorizing a real action.

The check also explains smaller contract differences. LangChain's absent nested
field in C01 matches the native-slot target even though its null artifact fails
the original null-sensitive T2 requirement. Microsoft's C01 native absence
matches too, while its added JSON text changes T2's text/structure relation.
Microsoft's H4 serialization also concatenates JSON in cases that this whole-text
parser cannot consume. These are explicit interface and parser-policy boundaries;
a different decoder might recover more. A strict-contract failure alone is not
proof that every downstream consumer loses the value.

\subsection{Where the differences arise}

The stage paths expose three recurring mechanisms. For structured values,
application carriers may retain an explicit object while model-facing carriers
retain only content-derived text. OpenAI keeps positive structured information
in application custom context and LangChain in a \texttt{ToolMessage} artifact;
neither auxiliary compartment receives credit for T2 at \stage{M3}. Microsoft
appends JSON text rather than preserving a separate structured state. Google
retains dictionary-valued function response fields through the measured lane.

Error state follows a different route. LangChain maps the distinction to message
status, and Google retains it in the result dictionary. OpenAI retains it in
application-side context but omits an actionable marker from the ordinary
model-client and captured-provider carriers. Microsoft represents the true case
through an error result at \stage{A2}/\stage{M3}, while \stage{H4} lacks the
marker required by T3.

Rich-content changes have different meanings across paths. OpenAI preserves
the required non-null atoms in six of its seven unmet rich fixtures but adds
null-valued optional fields; only C13 loses populated annotations and metadata.
Those are distinct mechanisms under the strict T4 signature. LangChain and
Microsoft still have unmet requirements under the optional-null policy, including
missing descriptors, annotations, metadata, or changed resource representation.
A data URL can preserve bytes and MIME without preserving all these atoms.
For a concrete type-and-identity loss, C10's source carried text beginning
\texttt{pwf-82254267465e} inside the resource at
\path{fixture://embedded/pwf-09975720945cc32bc2fe66be}.
Microsoft's \stage{A2}/\stage{M3} carriers retained that text as an ordinary text
item but not its resource kind or URI, so readable content survived while T4 was
unmet.

\section{Discussion}

\subsection{What post-wire verification should measure}

The central finding is that interoperability depends on an interface and a
consumer requirement together. A result can remain valid and readable while
losing the identity needed to retrieve a resource or the separate field expected
by a structured-value reader. An end-to-end success score can obscure where that
change occurred. The method supplies a narrower diagnostic: the public boundary
across which a required distinction ceases to be available. Its precision depends
on what the framework exposes. In three of the four paths, assuming a universal
five-stage pipeline would imply transformations or observations that do not exist.

Consumer requirements also determine which differences deserve attention. Byte
comparison can flag harmless encodings, while recursive value search can credit
information in a compartment the consumer never reads. Our optional-null analysis
shows the consequence of making this choice explicit: most OpenAI rich-content
failures disappear, even though the full-cohort divergence count stays the same.
The unchanged total conceals a substantial change in which paths satisfy the
contract. Comparisons should therefore report the semantic rule and the affected
cases alongside aggregate counts. A shared contract would make studies easier to
compare, but only if its requirements match the applications for which the
comparison is intended.

The oracle failure illustrates a further obligation for verification research:
the measurement instrument needs its own challenge cases. Passing the development
corpus had not established that the rich-content check compared everything its
contract required. A resource-text swap exposed that gap and changed the study's
conclusions after repair. Reporting the failed challenge, the general repair,
and both versions' cross-scores allows readers to assess what the validation
actually supports. This is more informative than the successful final score
alone, particularly for a small, authored suite.

\subsection{Consequences for framework and application developers}

The findings support the roadmap's effort to clarify result
handling~\cite{mcp-roadmap-2026}. For framework developers, documenting that an
adapter accepts MCP results is insufficient to describe its downstream behavior.
The interface contract should state which fields reach application code and the
model-facing path, including how errors and rich resources are represented.
That documentation can distinguish an intentional projection policy from a
regression. The corresponding tests can run locally at adapter boundaries;
provider calls are unnecessary for checking the transformations measured here.

Application developers face a related choice when adopting a structured-output
setting. The setting may improve JSON text without creating the field their
consumer expects. It can also change ordinary text handling, as the renderer
guards show. An upgrade test should therefore check both the desired structured
value and the existing consumer behavior that may change with it.

The conflicting-decision example gives this test a concrete purpose. OpenAI's
setting makes the text reader return the source structured decision; LangChain's
interceptor instead lets the reader identify ambiguity. Either behavior could
be useful, depending on whether the application requires channel precedence or
abstention. The complete readouts show why recovery counts alone cannot settle
that choice: JSON fallback recovers more values while also introducing wrong
and source-absent values. Applications should select a policy deliberately and
test its error behavior, rather than infer reliability from the number of values
successfully extracted. Whether these differences affect an LLM's decisions
requires a separate behavioral study.

\subsection{Threats to validity}

The consumer tasks are authored operational definitions, not an external gold
standard for every application. We reduce self-certification risk by grounding
atoms in immutable official schema/specification sources, separating primary
tasks from guards, preserving the first held-out miss, and using a fresh
post-repair suite. The external consumer audit motivates selected distinctions
but does not independently validate all task requirements. The 12-fault suites
remain small and purposively constructed. Their counts and operator coverage
cannot establish universal oracle accuracy.

Google's complete strict-contract pass record also raises representation-affinity
risk: exact authored contracts may favor a wire-like dictionary carrier. The
atoms were fixed against official sources before execution, and fresh fault
challenges tested semantic changes in framework-specific carriers. These checks
exposed an oracle miss; they do not establish that every consumer needs Google's
representation. The optional-null sensitivity makes one consequential contract
choice explicit without replacing the original results.

The exploratory replay uses known captures and a separately implemented decoder,
not independent authorship or external validation data. Gains are confined to
the original stress cases; other parsers or absence conventions could yield
different recoverability.

The cohort contains four purposively selected Python integrations at exact 2026
versions; releases, configurations, languages, or provider adapters can change
results. The 18 fixtures cover selected features but do not sample production
frequency, severity, or user impact. Counts describe the finite matrix, and stage
manifestations are dependent views rather than statistical replicates.

Microsoft documentation current after our study freeze describes retaining the
complete JSON-safe MCP result separately from the model-facing value for a Host
transport~\cite{microsoft-host-retention-2026}. That boundary is not the pinned
Python v1.17.0 path evaluated here. The newer guidance reinforces that these are
release- and transport-specific measurements, not durable framework rankings.

The local \stage{H4} transport establishes only the request body submitted by a
pinned provider SDK. We make no claim about a remote service, a model's effective
input, attention, generated behavior, or task quality. Some intermediate stages
are deliberately unobservable, and LangChain has no \stage{H4} in this cohort.
Those boundaries limit localization rather than justify private instrumentation.

The audit trail retains the five excluded runs and the accidental exposure of
their fragments described in Method. Inspection stopped immediately; corrections
before the admitted run addressed independently identified harness and projection
defects and integrity checks. The research question, fixtures, thresholds, and
tasks stayed frozen. Immutable destinations and full restarts preserve the history;
excluded outcomes contribute to no reported denominator.

\section{Conclusion}

Post-wire interoperability is a property of the interface a consumer uses and
the distinctions it requires. Our controlled study found changes in structured
values, error signals, and rich content across four pinned MCP integrations,
and localized them to the public boundaries the frameworks expose. Changing a
consumer policy can remove apparent failures or recover additional values, but
can also introduce different errors. These findings support regression tests
that specify the carrier, required semantics, and parsing policy together.
The method makes those requirements executable and their failures inspectable,
while the fault challenges demonstrate why the verifier itself must also be
tested. The resulting evidence concerns reproducible interface behavior;
production prevalence and effects on agent decisions remain separate questions.

\section*{Ethics and Reproducibility}

The study used benign synthetic fixtures, no human participants or personal
data, no remote model-provider request, and no paid service. The package
reconstructs analyses from sanitized normalized captures and supplies provenance,
locks, validation cases, deterministic code, and integrity checks. It does not
rerun framework environments or distribute private raw bundles.
The package is available at
\href{https://doi.org/10.5281/zenodo.22784143}{doi:10.5281/zenodo.22784143}.
The versioned archive is \texttt{se4agenticai-postwire-artifact-v4.zip}
(SHA-256: \path{fb374a8136ccd41e9ccdcff6df979cbbfcd98911eac5b4709aa1d92e6502f4b2}).

\section*{Acknowledgment}

AI (OpenAI Codex) was used for source discovery, implementation of the research
pipeline under the author's guidance, and manuscript drafting and language
refinement. The author substantially revised the manuscript and accepts full
responsibility for the article and reported results.

\IEEEtriggeratref{18}
\bibliographystyle{IEEEtran}
\bibliography{references}

\end{document}